\documentclass[12pt]{article}
\usepackage[dvips]{color}
\usepackage{amsmath}
\usepackage{amssymb}
\usepackage{graphicx}
\usepackage{epsfig}
\usepackage{cite}
\usepackage{subcaption}

\begin{document}

\title{\bf Observational Tests of the Generalized Uncertainty Principle: Shapiro Time Delay, Gravitational Redshift, and Geodetic Precession}
\author{{ \"{O}zg\"{u}r \"{O}kc\"{u} \thanks{Email: ozgur.okcu@ogr.iu.edu.tr}\hspace{1mm},
Ekrem Aydiner \thanks{Email: ekrem.aydiner@istanbul.edu.tr}} \\
{\small {\em Department of
		Physics, Faculty of Science, Istanbul University,
		}}\\{\small {\em Istanbul, 34134, Turkey}} }
\maketitle

\begin{abstract}
This paper is based on the study of the paper of Scardigli and Casadio [Eur. Phys. J. C (2015) 75:425] where the authors computed the light deflection and perihelion precession for the Generalized Uncertainty Principle (GUP) modified Schwarzschild metric. In the present work, we computed the gravitational tests such as Shapiro time delay, gravitational redshift, and geodetic precession for the GUP modified Schwarzschild metric. Using the results of Solar system experiments and observations, we obtain upper bounds for the GUP parameter $\beta$. Finally, we compare our bounds with other bounds in the literature.\\

\end{abstract}

\section{Introduction}
\label{intro}
Heisenberg uncertainty principle is the fundamental pillar of quantum mechanics, and its modification GUP may be unavoidable in the context of quantum gravity. The GUP serves as a useful tool against the inadequacies of general relativity (GR). Especially, GUP predicts the more sensible results of physics near the Planck length where the standard GR fails. Since GUP implies the minimal length at Planck scale, it removes the singularity predicted by standard GR.

Nowadays, GUP has been extensively studied in the literature since the papers based on string theory \cite{Veneziano1986,Amati1987,Amati1989,Gross1987,Gross1988,Konishi1990}. Based on Heisenberg microscope argument, Maggiore derived the GUP from a gedanken experiment \cite{Maggiore1993}. In Ref. \cite{Scardigli1999}, Scardigli  derived the GUP from a gedanken experiment involving micro black holes at the Planck scale.  Quantum mechanical applications of GUP can be found in Refs. \cite{Kempf1995,Kempf1994,Kempf1997,Nozari2012a,Chung2018}. On the other hand, GUP plays an important role on black hole thermodynamics. Since GUP introduces quantum gravity effects, modification of black hole thermodynamics may be necessary near the Planck scale \cite{Adler2001,Medved2004,Nozari2005,Nouicer2007,Nozari2008,Nowakowski2009,Arraut2009,Xiang2009,Jizba2010,Banerjee2010,Majumder2011,Scardigli2011,Nozari2012b,Gangopadhyay2014,Feng2016,Sakalli2016,Vagenas2017,Gecim2017,Sun2018,Gecim2018,Gecim2019,Kanzi2019,Okcu2020a,Gecim2020}. Moreover, GUP can be considered for various problems such as cosmological applications \cite{Awad2014,Salah2017,Okcu2020b}, particle accelerators \cite{Cavaglia2003,Cavaglia2004,Ali2012} and density of states applications \cite{Rama2001,Chang2002,Li2009,Vagenas2019}, etc\footnote{The reader may refer to review in Ref. \cite{Tawfik2014}}.

Besides the  theoretical investigations of GUP, there are studies which are devoted to measuring the upper bounds of various kinds of GUPs \cite{Das2008,Ali2011,Das2011,Pedram2011,Marin2013,Ghosh2014,Bawaj2015,Scardigli2015,Kohodadi2015,Gao2016,Gao2017,Feng2017,Kouwn2018,Bushev2019,Neves2020,Giardino2020}. It is normally assumed that GUP parameter $\beta$ is of the order of unity, but this assumption makes the GUP effects too small to be detectable. On the other hand, leaving this assumption gives the possibility for measuring the upper bounds of GUP parameter $\beta$ by using experiments and observations. Searching an upper bounds may be useful for implying the existence of  an intermediate length scale between electroweak and Planck length scales. The studies on this direction may also open a low energy window on phenomenology of quantum gravity since GUP effects can be considered various kinds of quantum mechanical systems.

In Ref. \cite{Scardigli2015}, Scardigli and Casadio computed the light deflection and perihelion precession for the GUP modified Schwarzschild metric\footnote{Various modified theories of gravity can be considered for Solar sytem tests. Using the observational results, researchers have recently found constrains on various modified theory of gravity \cite{Chakraborty2014,Ali2015,Lin2017,Bhattacharya2017,Gonzales2020,Bahamonde2020,Zhu2020}.}. First, they considered the GUP modified Hawking temperature of Schwarzschild black hole. Then, they obtained the corrections to the Schwarzschild metric from GUP modified Hawking temperature. Using the modified Schwarzschild metric, they obtained the GUP corrected light deflection and Perihelion precession. Comparing their theoretical results with precise astronomical measurements for both Solar system and binary pulsars, they obtained upper bounds of GUP parameter $\beta$. Moreover, their approach directly provides a novel method for the gravitational systems. Their approach can be applied other gravitational tests, and we can discuss new upper bounds of GUP parameter $\beta$. Therefore, we will study Shapiro time delay, gravitational redshift and geodetic precession for the GUP modified Schwarzschild metric.

The paper is arranged as follows. In Section (\ref{GUPSch}), we briefly review the GUP modified Schwarzschild metric in Ref. \cite{Scardigli2015}. In Section (\ref{PMGUP}), we  obtain the necessary equations to study Shapiro time delay and geodetic precession. At the end of Section (\ref{PMGUP}), we give the effective potential of particle around GUP modified Schwarzschild metric. In Section (\ref{TD}), we consider Shapiro time delay. In Section (\ref{GRS}), we study the gravitational redshift. In Section (\ref{GP}), we consider a spinning object in orbit around the GUP modified Schwarzschild metric. We obtain the components of its spin vector, and compute its geodesic precession. Finally, we discuss our results in Section (\ref{concl}).

\section{GUP-Modified Schwarzschild Metric}
\label{GUPSch}

In this section, we briefly review the GUP-modified Schwarzschild metric in Ref. \cite{Scardigli2015}. To obtain the GUP-modified Schwarzschild metric, we first consider the GUP-modified Hawking temperature. We start to consider the following GUP given as follows:
\begin{equation}
    \label{GUP}
    \Delta x\Delta p\geq\frac{1}{2}\left(1+4\beta G_{N}\Delta p^{2}\right)\,,
\end{equation}
where $\beta$ is the dimensionless parameter\footnote{We usually use the units $k_{B}=c=\hbar=1$, but we keep the physical constants during the numerical calculation of $\beta$ parameter.}. Considering the Eq.(\ref{GUP}) with standard dispersion relation $E=p$, we can write the wavelength of a photon
\begin{equation}
\label{deltaX}
\delta x\simeq\frac{1}{2E}+2\beta G_{N}E\,,
\end{equation}
where $E$ is the average energy of a photon. Taking the uncertainty of the photon wavelength is related with the Schwarzschild radius
\begin{equation}
\label{deltaX2}
\delta x\simeq 2\mu r_{S}=4G_{N}\mu M\,,
\end{equation}
and considering average energy of photon with Hawking temperature $E=T$, we obtain the mass-temperature relation from Eq.(\ref{deltaX})
\begin{equation}
\label{massTemperature}
4\mu G_{N}M\simeq\frac{1}{2T}+2\beta G_{N}T\,,
\end{equation}
where $\mu$ is the calibration factor fixed in the semiclassical limit, $\beta \rightarrow 0$. In the semiclassical limit, Hawking temperature is given by $T=\frac{1}{8\pi G_{N} M}$, and comparing the standard Hawking temperature with the temperature $T(\beta \rightarrow 0)$ in Eq. (\ref{massTemperature}), one can find $\mu=\pi$, so mass-temperature relation is given by
\begin{equation}
\label{massTemperature2}
M=\frac{1}{8\pi G_{N}T}+\beta\frac{T}{2\pi}\,,
\end{equation}
and modified temperature can be obtained from the above equation
\begin{equation}
\label{GUPtemp}
T=\frac{\pi}{\beta}\left(M-\sqrt{M^{2}-\frac{\beta}{4G_{N}\pi^{2}}}\right)\,.
\end{equation}

Now, we consider the simplest form of deformed Schwarzschild metric. The spherically symmetric metric is given by
\begin{equation}
\label{metric}
ds^{2}=-f(r)dt^{2}+\frac{dr^{2}}{f(r)}+r^{2}(d\theta^{2}+\sin^{2}\theta d\phi^{2})\,,
\end{equation}
\begin{equation}
\label{deformedMetric}
f(r)=1-\frac{2G_{N}M}{r}+\epsilon\frac{G_{N}^{2}M^{2}}{r^{2}}\,,
\end{equation}
where $\epsilon$ is a dimensionless parameter. From $f(r_{H}=0)$, the event horizon of deformed metric is given by
\begin{equation}
\label{eventH}
r_{H}=r_{S}\frac{1+\sqrt{1-\epsilon}}{2}\,,
\end{equation}
which is valid $\epsilon \leq 1$. Hence, the deformed Hawking temperature of metric in Eq.(\ref{deformedMetric}) is given by
\begin{equation}
\label{deformedTemp}
T(\epsilon)=\frac{f^{'}(r_{H})}{4\pi}=\frac{1}{2\pi G_{N}M}\frac{\sqrt{1-\epsilon}}{\left(1+\sqrt{1-\epsilon}\right)^{2}}\,,
\end{equation}
where prime denotes the derivative with respect to $r$. In order to relate the $\epsilon$ with $\beta$, deformed temperature in Eq. (\ref{deformedTemp}) must coincide with the modified temperature in Eq.(\ref{GUPtemp}), so the relation between $\epsilon$ and $\beta$ is given by
\begin{equation}
\label{relationBetaEpsilon}
\beta=-\frac{\pi^{2}G_{N}M^{2}}{\hbar c}\frac{\epsilon^{2}}{1-\epsilon}\,.
\end{equation}
The condition $\epsilon \leq 1$ clearly implies the negativity of GUP parameter $\beta$. So the deformed metric is able to define a GUP modified temperature for the negative $\beta$ parameter. Although the GUP parameter $\beta$ is usually defined positive in the literature, it is possible to define negative $\beta$ parameter. This situation is possible, when uncertainty relation is formulated on a crystal lattice \cite{Jizba2010}. This may imply the space time has granular or lattice structure at Planck scale.

\section{Particle Motion in GUP-Modified Schwarzschild Metric}
\label{PMGUP}

Let us start to consider the motion of the particle around the modified Schwarzschild metric in equatorial plane $\theta=\pi/2$. The Lagrangian of particle is given by
\begin{equation}
\label{Lagra}
\mathcal{L}=\frac{1}{2}g_{\mu\nu}\dot{x}^{\mu}\dot{x}^{\nu}=\frac{1}{2}\left[-f(r)\dot{t}^{2}+\frac{\dot{r}^{2}}{f(r)}+r^{2}\dot{\phi}^{2}\right] \,,
\end{equation}
where dot denotes the derivative with respect to affine parameter $\lambda$. Since the Lagrangian is independent of coordinates $t$ and $\phi$, the constants of motion can be obtained from generalized momentum $p_{\mu}$ of particle
\begin{equation}
\label{generalMoment}
p_{\mu}=\frac{\partial\mathcal{L}}{\partial\dot{x}^{\mu}}\,.
\end{equation}
From the above equation, one can easily obtain
\begin{equation}
\label{COM1}
p_{t}=\frac{\partial\mathcal{L}}{\partial\dot{t}}=-f(r)\dot{t}=-e\quad\Longrightarrow\quad\dot{t}=\frac{e}{f(r)}\,,
\end{equation}
\begin{equation}
\label{COM2}
p_{\phi}=\frac{\partial\mathcal{L}}{\partial\dot{\phi}}=r^{2}\dot{\phi}=\ensuremath{\ell}\quad\Longrightarrow\quad\dot{\phi}=\frac{\ell}{r^{2}}\,,
\end{equation}
where $e$ and $\ensuremath{\ell}$ are energy and angular momentum of the particle, respectively. Additionally, we have
\begin{equation}
\label{massCO}
g_{\mu\nu}\dot{x}^{\mu}\dot{x}^{\nu}=-k\,,
\end{equation}
where $k=1$ for massive particles and $k=0$ for massless particles. From Eqs. (\ref{COM1}), (\ref{COM2}) and (\ref{massCO}) we obtain
\begin{equation}
\label{intEq}
-\frac{e^{2}}{f(r)}+\frac{\dot{r}^{2}}{f(r)}+\frac{\ell^{2}}{r^{2}}=-k\,.
\end{equation}
Multiplying the Eq.(\ref{intEq}) by $f(r)/2$ and using $f(r)$ in Eq.(\ref{deformedMetric}), we write
\begin{equation}
\label{intEq2}
\frac{e^{2}-k}{2}=\frac{1}{2}\dot{r}^{2}+V_{eff}\,,
\end{equation}
where $V_{eff}$ is the effective potential of particle and is defined by
\begin{equation}
\label{Veff}
V_{eff}=-k\frac{G_{N}M}{r}+\frac{\ell^{2}+k\epsilon G_{N}^{2}M^{2}}{2r^{2}}-\frac{G_{N}M\ell^{2}}{r^{3}}+\epsilon\frac{G_{N}^{2}M^{2}\ell^{2}}{2r^{4}}\,.
\end{equation}
Later, we shall use extrema of the effective potential to obtain the stable circular orbit radius of the particle for geodetic precession.

\section{Shapiro Time Delay}
\label{TD}

In this section, we consider the time delay of electromagnetic signals for GUP deformed Schwarzschild metric. In 1964, Irwin Shapiro proposed a new test which is based on the measurement of photon time delay due to gravitational field \cite{Shapiro1964}. In order to calculate the time delay, we follow the argument of Ref. \cite{Bambi2018}. 

\begin{figure}
\label{STDFi}
\centerline{\includegraphics[width=7cm]{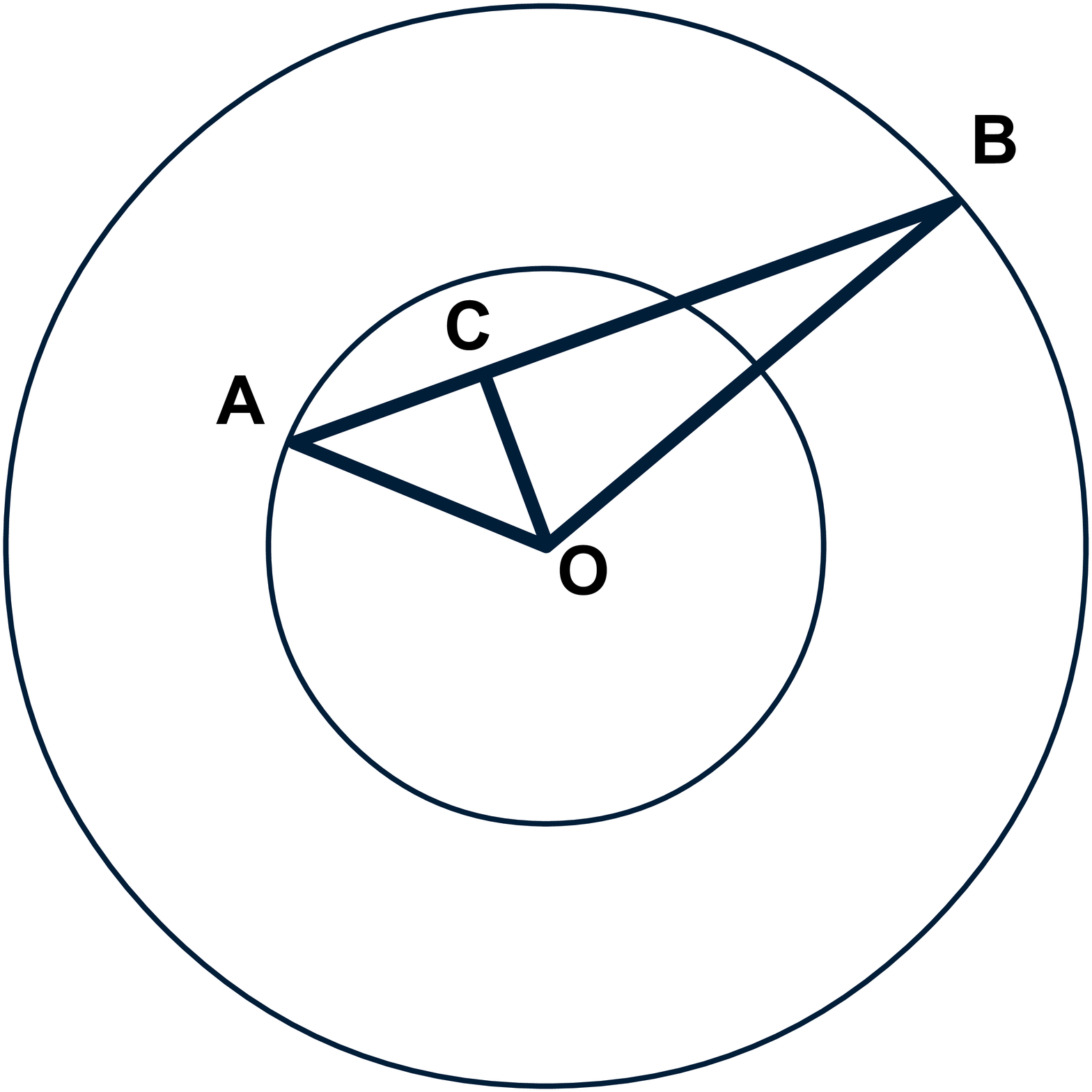}}
\vspace*{8pt}
\caption{Shapiro time delay. (See the text for more details.}
\end{figure}

The Fig. 1 is useful to describe the Shapiro time delay. We assume that Sun is located at point O. Let us suppose that electromagnetic signal is sent from point A with coordinates $(r_{A},\pi/2,\phi_{A})$ to point B with coordinates $(r_{B},\pi/2,\phi_{B})$. C is the closest point where the electromagnetic signal passes near the Sun. Now, we begin to calculate the travel time where electromagnetic signal is sent from A to B and reflected back to A in Solar system. With the help of
\begin{equation}
\label{DotR}
\frac{dr}{d\lambda}=\frac{dr}{dt}\frac{dt}{d\lambda}=\frac{dr}{dt}\frac{e}{f}\,,
\end{equation}
one can write the Eq. (\ref{intEq}) for massless particles ($k=0$)
\begin{equation}
\label{Mass0intEq}
\frac{e^{2}}{f(r)^{3}}\left(\frac{dr}{dt}\right)^{2}+\frac{\ell^{2}}{r^{2}}-\frac{e^{2}}{f(r)}=0\,.
\end{equation}
For the closest point $r=r_{C}$, we get $dr/dt=0$, and therefore
\begin{equation}
\label{angularMom}
\ell^{2}=\frac{e^{2}r_{C}^{2}}{f(r_{C})}\,.
\end{equation}
Using Eq. (\ref{angularMom}) in Eq. (\ref{Mass0intEq}), we obtain
\begin{equation}
\label{Mass0intEq2}
\left(\frac{dr}{dt}\right)^{2}=f(r)^{2}\left(1-\frac{f(r)r_{C}^{2}}{f(r_{C})r^{2}}\right)\,,
\end{equation}
or
\begin{equation}
\label{Mass0int3}
dt=\pm\frac{dr}{\sqrt{\left(1-\frac{f(r)r_{C}^{2}}{f(r_{C})r^{2}}\right)f(r)^{2}}}\,.
\end{equation}
Expanding in $r_{S}/r$ and $r_{S}/r_{C}$,
\begin{eqnarray}
\label{Series}
\frac{1}{\sqrt{\left(1-\frac{f(r)r_{C}^{2}}{f(r_{C})r^{2}}\right)f(r)^{2}}}	\approx	\frac{r}{\sqrt{r^{2}-r_{C}^{2}}}+\frac{r^{2}r_{S}}{\left(r^{2}-r_{C}^{2}\right)^{3/2}}+\frac{rr_{C}r_{S}}{2\left(r^{2}-r_{C}^{2}\right)^{3/2}}-\frac{3r_{C}^{2}r_{S}}{2\left(r^{2}-r_{C}^{2}\right)^{3/2}}\nonumber\\-\frac{3r_{C}^{3}r_{S}^{2}}{4\left(r^{2}-r_{C}^{2}\right)^{5/2}}-\frac{3r^{3}r_{S}^{2}(\epsilon-4)}{8(r^{2}-r_{C}^{2})^{5/2}}+\frac{3r_{C}^{2}r_{S}^{2}r(2\epsilon-7)}{8\left(r^{2}-r_{C}^{2}\right)^{5/2}}+\frac{3r_{C}^{4}r_{S}^{2}\left(5-\epsilon\right)}{8r\left(r^{2}-r_{C}^{2}\right)^{5/2}}\,.
\end{eqnarray}
The integrals of expanding terms are
\begin{equation}
\label{int1}
\int\frac{rdr}{\sqrt{r^{2}-r_{C}^{2}}}=\sqrt{r^{2}-r_{C}^{2}}\,,
\end{equation}
\begin{equation}
\label{int2}
\int\frac{r^{2}r_{S}dr}{\left(r^{2}-r_{C}^{2}\right)^{3/2}}=r_{S}\ln\left(r+\sqrt{r^{2}-r_{C}^{2}}\right)-\frac{rr_{s}}{\sqrt{r^{2}-r_{C}^{2}}}\,,
\end{equation}
\begin{equation}
\label{int3}
\int\frac{rr_{C}r_{S}dr}{2\left(r^{2}-r_{C}^{2}\right)^{3/2}}=\frac{-r_{C}r_{S}}{2\sqrt{r^{2}-r_{C}^{2}}}\,,
\end{equation}
\begin{equation}
\label{int4}
-\int\frac{3r_{C}^{2}r_{S}dr}{2\left(r^{2}-r_{C}^{2}\right)^{3/2}}=\frac{3rr_{S}}{2\sqrt{r^{2}-r_{C}^{2}}}\,,
\end{equation}
\begin{equation}
\label{int5}
-\int\frac{3r_{C}^{3}r_{S}^{2}dr}{4\left(r^{2}-r_{C}^{2}\right)^{5/2}}=\frac{r(3r_{C}^{2}-2r^{2})r_{S}^{2}}{4r_{O}\left(r^{2}-r_{C}^{2}\right)^{3/2}}\,,
\end{equation}
\begin{equation}
\label{int6}
-\int\frac{3r^{3}r_{S}^{2}(\epsilon-4)dr}{8\left(r^{2}-r_{C}^{2}\right)^{5/2}}=\frac{r_{S}^{2}\left(3r^{2}-2r_{C}^{2}\right)(\epsilon-4)}{8\left(r^{2}-r_{C}^{2}\right)^{3/2}}\,,
\end{equation}
\begin{equation}
\label{int7}
\int\frac{3r_{C}^{2}r_{S}^{2}r(2\epsilon-7)dr}{8\left(r^{2}-r_{C}^{2}\right)^{5/2}}=\frac{r_{C}^{2}r_{S}^{2}(7-2\epsilon)}{8\left(r^{2}-r_{C}^{2}\right)^{3/2}}\,,
\end{equation}
and
\begin{equation}
\label{int8}
\int\frac{3r_{C}^{4}r_{S}^{2}(5-\epsilon)dr}{8r\left(r^{2}-r_{C}^{2}\right)^{5/2}}=\frac{r_{S}^{2}(5-\epsilon)}{8r_{C}}\left[\frac{\left(3r^{2}-4r_{C}^{2}\right)r_{C}}{\left(r^{2}-r_{C}^{2}\right)^{3/2}}-3\arctan\left(\frac{r_{C}}{\sqrt{r^{2}-r_{C}^{2}}}\right)\right]\,.
\end{equation}
Using Eqs. (\ref{int1})-(\ref{int8}), the travel time of electromagnetic signal from point A to point C is given by
\begin{eqnarray}
\label{tAC}
&t_{AC}=\sqrt{r_{A}^{2}-r_{C}^{2}}+r_{S}\ln\left(\frac{r_{A}+\sqrt{r_{A}^{2}-r_{C}^{2}}}{C}\right)+\frac{r_{S}}{2}\sqrt{\frac{r_{A}-r_{C}}{r_{A}+r_{C}}}\left(1-\frac{r_{S}\left(4r_{A}+5r_{C}\right)}{4r_{C}\left(r_{A}+r_{C}\right)}\right)\nonumber\\&+\frac{3(\epsilon-5)r_{S}^{2}}{8r_{C}}\left[\arctan\left(\frac{r_{C}}{\sqrt{r_{A}^{2}-r_{C}^{2}}}\right)-\frac{\pi}{2}\right]\,.
\end{eqnarray}
Similarly, we find the travel time of electromagnetic signal from point C to point B
\begin{eqnarray}
\label{tBC}
&t_{BC}=\sqrt{r_{B}^{2}-r_{C}^{2}}+r_{S}\ln\left(\frac{r_{B}+\sqrt{r_{B}^{2}-r_{C}^{2}}}{r_{C}}\right)+\frac{r_{S}}{2}\sqrt{\frac{r_{B}-r_{C}}{r_{B}+r_{C}}}\left(1-\frac{r_{S}\left(4r_{B}+5r_{C}\right)}{4r_{C}\left(r_{B}+r_{C}\right)}\right)\nonumber\\&+\frac{3(\epsilon-5)r_{S}^{2}}{8r_{C}}\left[\arctan\left(\frac{r_{C}}{\sqrt{r_{B}^{2}-r_{C}^{2}}}\right)-\frac{\pi}{2}\right]\,.
\end{eqnarray}
The total travel time of electromagnetic signal is $t_{tot}=2t_{AC}+2t_{BC}$, and total travel time in flat spacetime is
\begin{equation}
\label{totFlat}
\widetilde{t}_{tot}=2\left(\sqrt{r_{A}^{2}-r_{C}^{2}}+\sqrt{r_{B}^{2}-r_{C}^{2}}\right)\,.
\end{equation}
Since $r_{C}\ll r_{A}, r_{B}$, the time delay is given by
\begin{equation}
\label{timeDelay}
\delta t=t_{tot}-\widetilde{t}_{tot}=4G_{N}M\left(1+\ln\left(\frac{4r_{A}r_{B}}{r_{C}^{2}}\right)-\frac{2G_{N}M}{r_{C}}\left(1+\frac{3\left(\epsilon-5\right)\pi}{8}\right)\right)\,.
\end{equation}

The time delay can be given in parameterized Post-Newtonian (PPN) formalism \cite{Will2014} 
\begin{equation}
\label{PPNTimeDelay}
\delta t=4G_{N}M\left(1+\left(\frac{1+\gamma}{2}\right)\ln\left(\frac{4r_{A}r_{B}}{r_{C}^{2}}\right)\right)\,,
\end{equation}
where $\gamma$ is a dimensionless PPN parameter. The time delay result of GR is recovered when $\gamma=1$. Comparing Eq. (\ref{timeDelay}) with Eq. (\ref{PPNTimeDelay}), we obtain
\begin{equation}
\label{Compare1}
|\gamma-1|=\frac{G_{N}M|15\pi-8-3\pi\epsilon|}{2c^{2}r_{C}\ln\left(\frac{4r_{A}r_{B}}{r_{C}^{2}}\right)}\,.
\end{equation}
Referring to measurement of the Cassini spacecraft \cite{Will2014,Bertotti2003}, the most stringent constraint of $\gamma$ parameter is $|\gamma-1|<2.3\times10^{-5}$. The Earth and spacecraft distances from Sun were $r_{A}=1$AB and $r_{B}=8.46$AB, respectively.  The closest distance of electromagnetic signal to Sun was only $r_{C}=1.6R_{\odot}$ where $R_{\odot}$  is the radius of Sun. Finally, we find
\begin{equation}
\label{TimeDelayBoundE}
-44.9<\epsilon<53.2\,,
\end{equation}
and with the constraint $\epsilon\leq1$
\begin{equation}
\label{TimeDelayBoundE2}
-44.9<\epsilon\leq1\,.
\end{equation}
If we employ the lower bound of Eq. (\ref{TimeDelayBoundE2}) in Eq. (\ref{relationBetaEpsilon}), we find the upper bound of GUP parameter $\beta$
\begin{equation}
\label{timeDelayBoundB}
|\beta|<3.6\times10^{78}\,.
\end{equation}
This bound is the same order of magnitude with the bound obtained from light deflection in Ref. \cite{Scardigli2015}. As it can be seen in Table 1, this bound is worse than the bounds obtained from quantum approaches. 

\section{Gravitational Redshift}
\label{GRS}
Let us consider that the electromagnetic signal moves from point $A$ to point $B$ in a gravitational field. We want to calculate the change of signal frequency for the deformed Schwarzschild metric. The gravitational redshift formula is given by
\begin{equation}
\label{RedShiftFormula}
\frac{\nu_{B}}{\nu_{A}}=\sqrt{\frac{f(r_{A})}{f(r_{B})}}\,,
\end{equation}
where $r_{A}$ and $r_{B}$ are the radial coordinates of the electromagnetic signal\footnote{Many textbooks on general relativity cover the gravitational redshift in detail. The reader may refer to Ref. \cite{Bambi2018}.}. In our case, the signal moves from the Earth surface $r_{A}=R_{\bigoplus}$ to height $h$, so we have $r_{B}=R_{\bigoplus}+h$. We can write the Eq. (\ref{RedShiftFormula}) as follows:
\begin{equation}
\label{RedShiftFormula2}
\frac{\nu_{B}}{\nu_{A}}=\sqrt{\frac{1-\frac{2G_{N}M}{r_{A}}+\epsilon\frac{G_{N}^{2}M^{2}}{r_{A}^{2}}}{1-\frac{2G_{N}M}{r_{B}}+\epsilon\frac{G_{N}^{2}M^{2}}{r_{B}^{2}}}}\,.
\end{equation}
Expanding Eq. (\ref{RedShiftFormula2}), the change of frequency can be obtained as
\begin{equation}
\label{RedShiftFormula3}
\frac{\Delta\nu}{\nu_{A}}=\frac{G_{N}M(r_{A}-r_{B})}{r_{A}r_{B}}\left[1+\frac{G_{N}M\left(\left(3-\epsilon\right)r_{A}+(1-\epsilon)r_{B}\right)}{2r_{A}r_{B}}\right]\,,
\end{equation}
where $\Delta\nu=\nu_{B}-\nu_{A}$. If we neglect the second term in the parentheses, Eq. (\ref{RedShiftFormula3}) gives the standard result of GR.  Referring to results of Pound-Snider experiment \cite{Pound1965}, relative deviation in the frequency from GR can be given by
\begin{equation}
\label{RSDeviation}
\frac{\frac{\Delta\nu}{\nu_{A}}-\left(\frac{\Delta\nu}{\nu_{A}}\right)^{GR}}{\left(\frac{\Delta\nu}{\nu_{A}}\right)^{GR}}<0.01\,.
\end{equation}
Employing (\ref{RedShiftFormula3}) in (\ref{RSDeviation}), we get
\begin{equation}
\label{RSDeviation2}
\frac{G_{N}M\left(\left(3-\epsilon\right)r_{A}+(1-\epsilon)r_{B}\right)}{2r_{A}r_{B}c^{2}}<0.01\,.
\end{equation}
The mass and radius of Earth are $M_{\bigoplus}=5.972\times10^{24}$kg and $R_{\bigoplus}=6378$km. The experiment was carried out in a tower with height $h=22.86$m. Therefore, we get
\begin{equation}
\label{GRSBoundE}
-1.4\times10^{7}<\epsilon\,,
\end{equation}
and Eq. (\ref{relationBetaEpsilon}) yields
\begin{equation}
\label{GRSBoundB}
|\beta|<1.1\times10^{73}\,.
\end{equation}
This bound is more stringent than the bound coming from Shapiro time delay, but it is worse than the bounds from quantum approaches.

\section{Geodetic Precession}
\label{GP}

Finally, we discuss the geodetic precession for the GUP modified metric solution. In this section, we follow the arguments of Ref. \cite{Hartle2014}. Let us begin to consider a gyroscope with a spin four-vector $\boldsymbol{s}$ in orbit around a spherical body of mass $M$. It is well known that a gyroscope with four-velocity $\boldsymbol{u}$ obeys the geodesic equation
\begin{equation}
\label{geodesicEq}
\frac{du^{\alpha}}{d\tau}+\Gamma_{\mu\nu}^{\alpha}u^{\mu}u^{\nu}=0\,,
\end{equation}
where $\Gamma_{\mu\nu}^{\alpha}$ is Christoffel symbol. In addition to geodesic equation, the motion of a gyroscope is described by
\begin{equation}
\label{gyroscopeEq}
\frac{ds^{\alpha}}{d\tau}+\Gamma_{\mu\nu}^{\alpha}s^{\mu}u^{\nu}=0\,.
\end{equation}
Eq. (\ref{gyroscopeEq}) is called gyroscope equation which describes the evolution of gyroscope spin. The spin and velocity four-vectors satisfy the condition
\begin{equation}
\label{orthogonalCond}
\boldsymbol{s}.\boldsymbol{u}=g_{\mu\nu}s^{\mu}u^{\nu}=0\,,
\end{equation}
and magnitude of spin $s_{*}$ is a constant of motion
\begin{equation}
\label{totalSpin}
\boldsymbol{s}.\boldsymbol{s}=g_{\mu\nu}s^{\mu}s^{\nu}=s_{*}^{2}\,.
\end{equation}
For simplicity, we consider circular orbit in equatorial plane, i.e., $\theta=\pi/2$, $\dot{r}=0=\dot{\theta}$. Therefore, the only spatial part of the four-velocity $\boldsymbol{u}$ is given by
\begin{equation}
\label{4veloSpat}  
u^{\phi}=\frac{d\phi}{d\tau}=\frac{d\phi}{dt}\frac{dt}{d\tau}=\Omega u^{t}\,,
\end{equation}
where $\Omega$ is the orbital angular velocity. The components of $\boldsymbol{u}$ are given by
\begin{equation}
\label{compOfFourVel}
\boldsymbol{u}=u^{t}(1,0,0,\Omega)\,.
\end{equation}
For the stable circular orbit in equatorial plane, Eq.(\ref{intEq2}) reduces to
\begin{equation}
\label{sCOE1}
\frac{e^{2}-1}{2}=V_{eff}\,,
\end{equation}
and radius $R$ of circular orbit is obtained from
\begin{equation}
\label{sCOE2}
\frac{dV_{eff}}{dr}=0\,.
\end{equation}
Neglecting the last term of effective potential $V_{eff}$ in Eq. (\ref{Veff}), one can find $R$ from Eq.(\ref{sCOE2}),
\begin{equation}
\label{RSCO}
R=\frac{\epsilon G_{N}^{2}M^{2}+\ell^{2}}{2G_{N}M}\left(1+\sqrt{1-\frac{12G_{N}^{2}M^{2}\ell^{2}}{\left(\epsilon G_{N}^{2}M^{2}+\ell^{2}\right)^{2}}}\right)\,.
\end{equation}
From Eq. (\ref{sCOE1}) and Eq. (\ref{RSCO}), $e^{2}$ and $\ell^{2}$ are given by
\begin{equation}
\label{e2}
e^{2}=\left(1-\frac{2G_{N}M}{R}+\frac{\epsilon G_{N}^{2}M^{2}}{R^{2}}\right)\left(1+\frac{\ell^{2}}{R^{2}}\right)\,,
\end{equation}
\begin{equation}
\label{l2}
\ell^{2}=G_{N}MR\left(1-\frac{\epsilon G_{N}M}{R}\right)\left(1-\frac{3G_{N}M}{R}\right)^{-1}\,,
\end{equation}
respectively. Using Eqs. (\ref{COM1}) and (\ref{COM2}), the angular orbital velocity is given by
\begin{equation}
\label{AOV}
\Omega=\frac{d\phi}{dt}=\frac{d\phi}{d\tau}\frac{d\tau}{dt}=\frac{f(r)}{r^{2}}\frac{\ell}{e}\,.
\end{equation}
Inserting Eqs. (\ref{e2}) and (\ref{l2}) in Eq. (\ref{AOV}), we have
\begin{equation}
\label{AOV2}
\Omega^{2}=\frac{G_{N}M}{R^{3}}\left(1-\frac{\epsilon G_{N}M}{R}\right)\left(1-\frac{2G_{N}M}{R}+\frac{\epsilon G_{N}^{2}M^{2}}{R^{2}}\right)\left(1-\frac{2G_{N}M}{R}-\frac{\epsilon G_{N}^{2}M^{2}}{R^{2}}\right)^{-1}\,,
\end{equation}
or
\begin{equation}
\label{AOV3}
\Omega^{2}=\frac{G_{N}M}{R^{3}}\left(1-\frac{\epsilon G_{N}M}{R}\right)\left(1-\frac{4G_{N}^{2}M^{2}}{R^{2}}+\frac{2\epsilon G_{N}^{2}M^{2}}{R^{2}}+\frac{\epsilon^{2}G_{N}^{4}M^{4}}{R^{4}}\right)\,,
\end{equation}
where we use $\left(1+\alpha\right)^{n}\approx1+n\alpha$ for $\alpha\ll1$ in Eq. (\ref{AOV2}). Neglecting the higher order terms, we can finally write
\begin{equation}
\label{AOV4}
\Omega=\sqrt{\frac{G_{N}M}{R^{3}}\left(1-\frac{\epsilon G_{N}M}{R}\right)}\,.
\end{equation}
Now, we can solve the gyroscope equation in Eq. (\ref{gyroscopeEq}). We assume that spin vector is initially radial directed, i.e., $s^{t}(0)=s^{\theta}(0)=s^{\phi}(0)=0$. Using the condition in Eq. (\ref{orthogonalCond}) we get the relation between $s^{t}$ and $s^{\phi}$
\begin{equation}
\label{Relation(ST)(SPhi)}
s^{t}=\Omega R^{2}\left(1-\frac{2G_{N}M}{R}+\frac{\epsilon G_{N}^{2}M^{2}}{R^{2}}\right)^{-1}s^{\phi}\,.
\end{equation}
With the help of Eqs. (\ref{compOfFourVel}) and (\ref{Relation(ST)(SPhi)}), the radial component of gyroscope equation is given by
\begin{equation}
\label{gyroRadial}
\frac{ds^{r}}{d\tau}+\left(3G_{N}M-R-\frac{2\epsilon G_{N}^{2}M^{2}}{R}\right)\Omega s^{\phi}u^{t}=0\,.
\end{equation}
The $\theta$ and $\phi$ components of gyroscope equation are given by
\begin{equation}
\label{gyroTheta}
\frac{ds^{\theta}}{d\tau}-\sin\theta\cos\theta s^{\phi}u^{\phi}=0\,,
\end{equation}
\begin{equation}
\label{gyroPhi}
\frac{ds^{\phi}}{d\tau}+\frac{\Omega}{R}s^{r}u^{t}=0
\end{equation}
Since the trajectory is in the equatorial plane, Eq.(\ref{gyroTheta}) reduces to $\frac{ds^{\theta}}{d\tau}=0$. Imposing the initial condition $s^{\theta}(0)=0$, one finds $s^{\theta}$ remains zero throughout trajectory. Using $u^{t}=dt/d\tau$ in Eqs. (\ref{gyroRadial}) and (\ref{gyroPhi}), we have
\begin{equation}
\label{gyroRadial2}
\frac{ds^{r}}{dt}+\left(3G_{N}M-R-\frac{2\epsilon G_{N}^{2}M^{2}}{R}\right)\Omega s^{\phi}=0\,,
\end{equation}
\begin{equation}
\label{gyroPhi2}
\frac{ds^{\phi}}{dt}+\frac{\Omega}{R}s^{r}=0\,.
\end{equation}
Employing Eq. (\ref{gyroPhi2}) in Eq. (\ref{gyroRadial2}) gives
\begin{equation}
\label{SHOLE}
\frac{d^{2}s^{\phi}}{dt^{2}}+\tilde{\Omega}^{2}s^{\phi}=0\,,
\end{equation}
where we define
\begin{equation}
\label{tildeOmega}
\tilde{\Omega}=\sqrt{1-\frac{3G_{N}M}{R}+\frac{2\epsilon G_{N}^{2}M^{2}}{R^{2}}}\Omega\,.
\end{equation}
Eqs. (\ref{gyroRadial2}) and (\ref{SHOLE}) yield the solutions
\begin{equation}
\label{sol1}
s^{r}(t)=s_{*}\sqrt{1-\frac{2G_{N}M}{R}+\frac{\epsilon G_{N}^{2}M^{2}}{R^{2}}}\cos\left(\tilde{\Omega}t\right)\,,
\end{equation}
\begin{equation}
\label{sol1}
s^{\phi}(t)=-s_{*}\frac{\Omega}{\tilde{\Omega}R}\sqrt{1-\frac{2G_{N}M}{R}+\frac{\epsilon G_{N}^{2}M^{2}}{R^{2}}}\sin\left(\tilde{\Omega}t\right)\,,
\end{equation}
where we use the conditions $\boldsymbol{s}.\boldsymbol{s}=s_{*}^{2}$ and $s^{t}(0)=s^{\phi}(0)=0$.

The spin starts along a unit radial vector $\boldsymbol{e_{\hat{r}}}$ with components $\left(0,\sqrt{f(r)},0,0\right)$. If one rotation along circular orbit takes time $P=2\pi/\Omega$,  we can find the change of spin direction,
\begin{equation}
\label{geodetic1}
\left[\frac{\boldsymbol{s}}{s_{*}}.\boldsymbol{e}_{\hat{r}}\right]_{t=P}=\cos\left(\frac{2\pi\tilde{\Omega}}{\Omega}\right)\,.
\end{equation}
As a result, we find
\begin{equation}
\label{geodetic2}
\Delta\Phi_{geodetic}=2\pi-2\pi\sqrt{1-\frac{3G_{N}M}{R}+\frac{2\epsilon G_{N}^{2}M^{2}}{R^{2}}}\,.
\end{equation}
For the Solar system, the geodetic precession is approximately given by
\begin{equation}
\label{geodetic3}
\Delta\Phi_{geodetic}=\Delta\Phi_{GR}\left(1-\frac{2\epsilon G_{N}M}{3Rc^{2}}\right)\,,
\end{equation}
where GR result is $\Delta\Phi_{GR}=\frac{3\pi G_{N}M}{Rc^{2}}$. Referring to measurements of Gravity Probe B (GPB) \cite{Everitt2011}, we can find an upper bound for $\beta$. Geodetic precession was measured by GPB
\begin{equation}
\label{GPB}
\Delta\Phi_{geodetic}=\left(6601.8\pm18.3\right)mas/year\,.
\end{equation}
Considering the GPB at an altitude of $642$km and with an orbital period of $97.65$min, GR predicts  $\Delta\Phi_{GR}=6606.1$mass/year. Therefore, we get
\begin{equation}
\label{EpsilonGeodeticBound}
-5\times10^{6}<\epsilon<8.1\times10^{6}\,,
\end{equation}
and with the constraint $\epsilon\leq1$
\begin{equation}
\label{EpsilonGeodeticBound2}
-5\times10^{6}<\epsilon\leq 1\,.
\end{equation}
Using Eq. (\ref{relationBetaEpsilon}), we finally obtain
\begin{equation}
\label{BetaGeodeticBound}
|\beta|<3.7\times10^{72}\,.
\end{equation}
This bound is the most stringent bound in this paper, but it is looser than the bounds from quantum experiments.

\section{Discussions and Conclusions}
\label{concl}

This work is based on the paper of Scardigli and Casadio \cite{Scardigli2015} where the authors deformed the Schwarzschild metric to reproduce GUP modified Hawking temperature, and then they computed light deflection and perihelion precession for the deformed metric. They compared their theoretical results with astronomical measurements. They finally obtained the upper bounds of parameter $\beta$. In this work, we extended their approach to gravitational tests such as Shapiro time delay, gravitational redshift and geodetic precession. 

%%%%%%%%%%%%%%%%%%%%%%%%%TABLE%%%%%%%%%%%%%%%%%%%%%%%%%%%%%%%%%%%%%
\begin{table}[ht]
	\caption{Upper bounds of GUP $\beta$ obtained from various experiments.} \label{table1} % title of Table
	\centering % used for centering table
	\begin{tabular}{ c c c } % centered columns (4 columns)
		\hline\hline 
	    Experiment & $\beta$ & Reference \\ [0.5ex] % inserts table %heading
		\hline % inserts single horizontal line
		Lamb shift  &  $10^{36}$ & \cite{Das2008}  \\ % inserting body of the table
		Landau levels & $10^{50}$ &  \cite{Das2008}  \\
		Scanning tunneling microscope & $10^{21}$ & \cite{Das2008}  \\
		Harmonic oscillators  & $10^{6}$ & \cite{Bushev2019}  \\
		Gravitational waves  &  $10^{60}$ & \cite{Feng2017}  \\ 
		Light deflection    &  $10^{78}$  & \cite{Scardigli2015}  \\
		Perihelion precession &  $10^{69}$ &  \cite{Scardigli2015} \\
		Pulsar periastron shift &  $10^{71}$ &  \cite{Scardigli2015} \\
		Black hole shadow  &   $10^{90}$ & \cite{Neves2020}   \\
		Cosmological constraints  &  $10^{81}$  & \cite{Kouwn2018} \\
		Cosmological constraints &  $10^{59}$,$10^{81}$ &  \cite{Giardino2020} \\
		\textbf{Shapiro time delay}  & $10^{78}$ &  in this study \\ 
		\textbf{Gravitational red-shift} & $10^{73}$  &  in this study \\ 
		\textbf{Geodetic precession}  & $10^{72}$	&  in this study \\      [1ex] % [1ex] adds vertical space
		\hline %inserts single line
	\end{tabular}
	\label{Table1} % is used to refer this table in the text
\end{table}
%%%%%%%%%%%%%%%%%%%%%%%%%%%%%%%%%%%%%%%%%%%%%%%%%%%%%%%%%%%%%%%%%%%%

In Table 1, we give the upper bounds of GUP parameter $\beta$ from experiments. As we can see in Table 1, quantum experiments provide  more stringent bounds. Unlike the quantum experiments, gravitational tests give looser bounds. In Ref. \cite{Neves2020},  upper bound $\beta < 10^{90}$ was obtained from black hole shadow. To the best of our knowledge, this bound has the worst value. On the other hand, authors of Refs. \cite{Kouwn2018,Giardino2020} obtained the more stringent bounds by using GUP-modified Friedmann equations with cosmological constraints. Based on the GUP deformation of dispersion relation, authors of Ref. \cite{Feng2017} obtained $\beta < 10^{60}$ from the gravitational wave event GW150914 \cite{Ligo2016,Ligo2016b}.

The bounds in this work are not tighter than the bounds from quantum experiments.  In addition to results of Ref. \cite{Scardigli2015}, namely light deflection ($|\beta|<10^{78}$), perihelion precession of Mercury ($|\beta|<10^{69}$) and pulsar periastron shift ($|\beta|<10^{71}$), we investigated Shapiro time delay ($|\beta|<10^{78}$), gravitational redshift ($|\beta|<10^{73}$) and geodetic precession ($|\beta|<10^{72}$). The bound $\beta < 10^{72}$ from geodetic precession is the most stringent bound in this work, but it is clearly worse than the bounds from quantum approaches. Comparing our bounds with the bounds in Ref.\cite{Scardigli2015}, the bound $|\beta|<10^{69}$ from perihelion precession is  the most stringent bound.  Apart from Ref. \cite{Scardigli2015} and this paper, the author of Ref. \cite{Ghosh2014}  reported the bounds $\beta < 10^{26}$ from gravitational redshift, $\beta < 10^{19}$ from the law of reciprocal actions, $\beta < 10^{19}$ from universality of free fall. These bounds are clearly more stringent than the other gravitational bounds. The method in Ref. \cite{Ghosh2014} is based on the deformed Poisson brackets which leads to the violation of equivalence principle (EP). In a recent paper \cite{Casadio2020}, authors showed that the deformation of Poisson brackets has some defects such as  huge violation of EP for astronomical objects, badly defined classical limit, etc. For example, the trajectory of test particle in Ref. \cite{Casadio2020} is given by
\begin{eqnarray}
\ddot{r}\backsimeq-\frac{G_{N}M}{r^{2}}\left(1+4\beta\frac{m^{2}}{m_{p}^{2}}\dot{r}^{2}\right),
\end{eqnarray}
which clearly depends on its mass $m$ and velocity $\dot{r}$. This implies the violation of EP. Furthermore, deformed term increases quadratically with the mass of test particle. This may lead to huge deviation from GR \footnote{For the more details, the reader may refer to Ref. \cite{Scardigli2019}, appendices in Ref. \cite{Scardigli2015}, and Eqs. (10), (14), (15) in Ref. \cite{Casadio2020}.}. Another problem is the divergent of the commutator  due to badly defined classical limit \footnote{See Eqs. (19) and (21) in Ref. \cite{Casadio2020}.}.  On the other hand, deformed metric in  Eq. (\ref{deformedMetric}) is only related to GUP modified temperature without EP violation.  Since Poisson brackets are not deformed, the above mentioned defects are not avaliable for deformed metric.

Measuring the upper bounds of GUP parameter $\beta$ may provide us to consider the phenomenology of quantum gravity beyond the Planck scale. Gravitational constraints on $\beta$ may open a large structure windows on the phenomenology of quantum gravity. We hope to report in future studies.

\section*{Acknowledgement}
The authors thank the anonymous reviewer for his/her helpful and constructive comments. \"{O}zg\"{u}r \"{O}kc\"{u} thanks Can Onur Keser for drawing Fig. 1.

\end{document}